\documentclass[12pt]{article}
\usepackage{graphicx}

\newcommand{\Frac}[2]{\frac{\displaystyle #1}{\displaystyle #2}}
\newcommand{\beq}{\begin{equation}}
\newcommand{\eeq}{\end{equation}}
\newcommand{\beqn}{\begin{eqnarray}}
\newcommand{\eeqn}{\end{eqnarray}}
\newcommand{\beqns}{\begin{eqnarray*}}
\newcommand{\eeqns}{\end{eqnarray*}}

\begin{document}
\begin{titlepage}
\begin{center}

\hfill USTC-ICTS-03-15\\
\hfill  December 2003

\vspace{2.5cm}

{\Large {\bf   On the electron-positron energy  asymmetry in
$K_L\to\pi^0 e^+e^-$
\\}} \vspace*{1.0cm}
 {  Dao-Neng Gao$^\dagger$} \vspace*{0.3cm} \\
{\it\small Interdisciplinary Center for Theoretical Study and
Department of Modern Physics,\\ University of Science and
Technology of China, Hefei, Anhui 230026 China}

\vspace*{1cm}
\end{center}
\begin{abstract}
\noindent Motivated by some new experimental and theoretical
efforts, we update the theoretical analysis of the
electron-positron energy asymmetry, which arises, in
$K_{L}\to\pi^0 e^+e^-$, from the two-photon intermediate state in
the standard model. It is found that the measurement of this
asymmetry in future experiments may increase our understanding of
the $K_{L,~S}\to \pi^0 e^+ e^-$ decays, which would thus provide
some useful information on quark flavor physics. Meanwhile, in the
standard model the electron-positron energy asymmetry in the decay
of $K_S\to\pi^0 e^+e^-$  is expected to be vanishingly small,
therefore the asymmetry in $K_S\to\pi^0 e^+ e^-$ might be a very
interesting quantity to explore new physics scenarios.
\end{abstract}

\vfill
\noindent

$^{\dagger}$ E-mail:~gaodn@ustc.edu.cn
\end{titlepage}

The flavor-changing neutral-current process $K_{L}\to\pi^0 e^+
e^-$ has been recognized as one of the most interesting rare kaon
decays for a long time \cite{BK00, DI98}. It is known that the
decay rate of the transition is given by a sum of comparable
CP-conserving, direct and indirect CP-violating contributions
\cite{EPR88, DG95}, which in general leads to some difficulties in
carrying out an accuracy analysis of the decay. Very recently, the
authors of Ref. \cite{BDI03}, based on two new experimental
results from NA48 Collaboration: the first observation of
$K_S\to\pi^0 e^+e^-$ \cite{NA4803} and the precise measurement of
the $K_L\to\pi^0\gamma\gamma$ spectrum \cite{NA4802}, have argued
that the CP-conserving part is essentially negligible for the rate
of $K_L\to\pi^0 e^+e^-$, and the standard model (SM) contribution
to its branching ratio has been predicted as $BR(K_L\to\pi^0
e^+e^-)_{\rm SM}\simeq 3\times 10^{-11}$, which is dominated by
CP-violating components with about 40\% from the direct
CP-violating amplitude, through the interference with the indirect
CP-violating one. Recalling the new experimental upper bound
$BR(K_L\to\pi^0 e^+e^-)<2.8\times 10^{-10}$ (90\% C.L.)
\cite{KTeV03}, observation of the decay rate substantially higher
than the SM expectation would therefore signal new physics. On the
other hand, in order to deeply understand this decay mode, many
interesting observables besides the decay rate of $K_L\to\pi^0
\ell^+\ell^-$ ($\ell=e$ and $\mu$), such as the muon polarization
effects and the lepton energy asymmetries, have been studied in
the past literature \cite{DG95, LMS88, HS93, DM01, DMT02}.
Motivated by the new experimental and theoretical efforts
mentioned above, in this paper we will update the theoretical
analysis of the electron-positron energy asymmetry in the decay of
$K_{L}\to\pi^0 e^+e^-$.

The general invariant amplitude for
$K_{L}(p)\to\pi^0(p_\pi)e^+(p_+)e^-(p_-)$ in terms of the scalar,
pseudoscalar, vector, and axial-vector form factors, denoted by
$F_S$, $F_P$, $F_V$, and $F_A$,  can be parameterized as
\cite{ANBG91, CQ03} \beq\label{IA1} {\cal M}=F_S\bar{e} e+i
F_P\bar{e}\gamma_5 e+F_Vp^\mu\bar{e}\gamma_\mu e+
F_Ap^\mu\bar{e}\gamma_\mu\gamma_5 e,\eeq where $p$, $p_\pi$, and
$p_\pm$ are the four-momenta of $K_{L}$, $\pi^0$, and $e^\pm$,
respectively. The differential decay rate in the $K_L$ rest frame
is \beqn\label{rate0}\frac{d\Gamma}{dE_+dE_-}&=&\frac{1}{2^4
\pi^3}\left[|F_S|^2 \frac{1}{2}(s-4 m_e^2)+|F_P|^2 \frac{1}{2}s
+|F_V|^2m_K^2\left(2 E_+ E_--s/2\right)\right. \nonumber\\
&&+|F_A|^2 m_K^2\left(2 E_+ E_--s/2+2
m_e^2\right)+2 {\rm Re}(F_S^* F_V)m_e m_K (E_+-E_-)\nonumber\\
&&\left. +{\rm Im}(F_P F_A^*)m_e(m_\pi^2-m_K^2-s)\right], \eeqn
where $s=(p_++p_-)^2$ and $E_\pm$ denotes the energy of $e^\pm$ in
the $K_L$ rest frame. One can find that, the terms which are
antisymmetric under the exchange of $E_+\leftrightarrow E_-$ in
Eq. (\ref{rate0}) will lead to an asymmetry in the
electron-positron energy distribution \beq{\cal
A}=\frac{N(E_+>E_-)-N(E_+<E_-)}{N(E_+>E_-)+N(E_+<E_-)}. \eeq It is
convenient to introduce two dimensionless variables, $z=s/m_K^2$,
and $\theta$, the angle between the three-momentum of the kaon and
the three-momentum of the $e^-$ in the dilepton rest frame, to
rewrite Eq. (\ref{rate0}) as \cite{CQ03} \beqn\label{rate1}
\frac{d\Gamma}{dz d\cos\theta}=\frac{m_K^5\beta
\lambda^{1/2}(1,z,r_\pi^2)}{2^8\pi^3}\left\{\left|\frac{F_S}{m_K}\right|^2
z \beta^2+\left|\frac{F_P}{m_K}\right|^2
z+|F_V|^2\frac{1}{4}\lambda(1,z,r_\pi^2)(1-\beta^2
\cos^2\theta)\right.\nonumber \\
\left. +|F_A|^2\left[\frac{1}{4}\lambda(1,z,r_\pi^2)(1-\beta^2
\cos^2\theta)+4 r_e^2\right]\right.\nonumber\\\left.+\frac{{\rm
Re}(F_S F_V^*)}{m_K}
2r_e~\beta\lambda^{1/2}(1,z,r_\pi^2)\cos\theta +\frac{{\rm
Im}(F_PF_A^*)}{m_K}2r_e (r_\pi^2-1-z) \right\},
 \eeqn
 where $\lambda(a,b,c)=a^2+b^2+c^2-2(ab+ac+bc)$,
 $r_e=m_e/m_K$, $r_\pi=m_\pi/m_K$,
 $\beta=\sqrt{1-4r_e^2/z}$, and we have used
 \beq\label{epm}
 E_\pm=\frac{m_K}{4}\left[(1-r_\pi^2+z)\mp \beta \lambda^{1/2}
 (1,z,r_\pi^2)\cos\theta\right]. \eeq
 Thus the differential electron-positron energy asymmetry can be defined as
 \beq\label{AFB1}
 {\cal A}(z)=\frac{d\Gamma(E_+>E_-)/dz-d\Gamma(E_+<E_-)/dz}{d\Gamma(E_+>E_-)/dz
 +d\Gamma(E_+<E_-)/dz},
 \eeq
with \beqn \frac{d\Gamma}{dz}(E_+>E_-)= \int^1_0
\frac{d\Gamma}{dzd\cos\theta}~d\cos\theta,\nonumber \\ \\
\frac{d\Gamma}{dz}(E_+<E_-)=\int^0_{-1}
 \frac{d\Gamma}{dzd\cos\theta}~d\cos\theta\nonumber,
 \eeqn
 and the phase space in terms of  $z$ and $\cos\theta$ is given by
\beq\label{phasespace}
 4r_e^2\le
 z\le(1-r_\pi)^2,\;\;\;\;\;-1\le \cos\theta\le 1.\eeq

It is easy to see from Eqs. (\ref{rate1}) and (\ref{AFB1}) that,
in general only two form factors $F_S$ and $F_V$ will play
relevant roles in obtaining the significant asymmetry. In the SM,
the vector form factor $F_V$ has both CP-conserving and
CP-violating contributions; while the scalar form factor $F_S$
gets the contribution only from the two-photon intermediate state
via the transition $K_L\to\pi^0\gamma^*\gamma^*\to \pi^0 e^+ e^-$.
The decays $K_S\to\pi^0\ell^+\ell^-$, which are related to the
indirect CP-violating component of $K_L\to\pi^0 e^+ e^-$ and
dominated by low energy dynamics, have been studied in chiral
perturbation theory beyond the leading order, and the
corresponding vector form factor is parameterized as \cite{DEIP98}
\beq\label{FVS1} F_V^S=-\frac{\alpha}{2\pi m_K^2}\left[G_F
m_K^2(a_S+b_S z)+W^{\pi\pi}_S(z)\right], \eeq where the real
parameters $a_S$ and $b_S$ encode local contributions up to
$O(p^6)$ in chiral expansion, and the non-analytic function
$W_S^{\pi\pi}(z)$, which is generated by the $\pi\pi$ loop and can
be completely determined in terms of the physical $K\to3\pi$
amplitude, is known to be very small due to the $\Delta I=3/2$
suppression of the $K_S\to\pi^+\pi^-\pi^0$ amplitude
\cite{DEIP98}. Therefore as a good approximation, Eq. (\ref{FVS1})
can be simplified as \beq\label{FVS}F_V^S=-\frac{\alpha
G_F}{2\pi}(a_S+b_S z).\eeq Motivated by vector meson dominance
(VMD), we have \cite{DEIP98} \beq\label{VMD}
b_S=\frac{a_S}{r_V^2},\eeq where $r_V=m_V/m_K$, and $m_V$ is the
vector meson mass. The first experimental evidence of $K_S\to\pi^0
e^+e^-$ has been reported by NA48 Collaboration \cite{NA4803}:
\beq\label{KSrate}{\rm Br}(K_S\to\pi^0 e^+e^-)_{m_{ee}>165{\rm
MeV}}=(3.0^{+1.5}_{-1.2}\pm0.2)\times 10^{-9},\eeq which leads to
the determination of $a_S$ \cite{NA4803}
\beq\label{aS}|a_S|=1.06^{+0.26}_{-0.21}~.\eeq

The direct CP-violating amplitude for $K_L\to\pi^0 e^+e^-$ is
dominated by short-distance dynamics and is calculable with high
accuracy in perturbation theory \cite{BDI03, BBL96, BLMM94}.  Thus
the CP-violating vector form factor including indirect and direct
CP-violating parts can be written as  \beq\label{FVL1}
(F_V^L)_{\rm CPV}=-\frac{\alpha
G_F}{2\pi}\left(1+\frac{z}{r_V^2}\right) \left[ a_S|\epsilon|
e^{i\phi_\epsilon}-i\frac{4\pi y_{7V}(\mu)}{\sqrt{2}\alpha}{\rm
Im}\lambda_t\right],\eeq where $|\epsilon|=(2.28\pm 0.02)\times
10^{-3}$ and $\phi_\epsilon=43.5^o$ \cite{PDG} encode the indirect
CP-violating contribution due to $K^0-\overline{K}^0$ mixing, and
the $y_{7V}$ part in Eq. (\ref{FVL1}) is from the short-distance
direct CP violation, $y_{7V}=(0.073\pm 0.04)\alpha (M_Z)$  with
$\alpha (M_Z)=1/129$ \cite{BBL96}, and $\lambda_t=V_{ts}^*V_{td}$
with ${\rm Im}\lambda_t=(1.36\pm 0.12)\times 10^{-4}$
\cite{CKM03}. Note that the VMD relation [Eq. (\ref{VMD})] has
been taken into account in deriving Eq. (\ref{FVL1}).

The CP-conserving amplitude for $K_L\to\pi^0 e^+e^-$ is generated
by the long-distance transition
$K_L\to\pi^0\gamma^*\gamma^*\to\pi^0 e^+e^-$, which can contribute
to both  $F_S$ and $F_V$. The general invariant amplitude for the
$K_L(p)\to\pi^0(p_\pi)\gamma(q_1,\epsilon_1)\gamma(q_2,
\epsilon_2)$ decay is given by \beqn\label{IA2gamma1}{\cal
A}(K_L\to\pi^0\gamma\gamma)&=&\frac{G_8\alpha}{4\pi}\epsilon_{1\mu}(q_1)\epsilon_{2\nu}(q_2)[A(y,z)(q_2^\mu
q_1^\nu-q_1\cdot q_2 g^{\mu\nu})\nonumber\\
&+& \frac{2B(y,z)}{m_K^2}(p\cdot q_1 q_2^\mu p^\nu+p\cdot q_2
p^\mu q_1^\nu -p\cdot q_1 p\cdot q_2 g^{\mu\nu}-q_1\cdot q_2 p^\mu
p^\nu)],\eeqn  where $y=p\cdot(q_1-q_2)/m_K^2$, $
z=(q_1+q_2)^2/m_K^2$, and $|G_8|=9.2\times 10^{-6}$ GeV$^{-2}$.
Within the framework of chiral perturbation theory, the amplitude
$A(y,z)$ will receive non-vanishing contribution at $O(p^4)$
\cite{EPR87}; while the leading order contribution to $B(y,z)$
starts from $O(p^6)$, which has been extensively studied by
including both the unitarity corrections and the local terms
generated by vector resonance exchange \cite{EPR90, CDM93, CEP93,
KH94, DP97, GV01}. It is known that, via the transition
$K_L\to\pi^0\gamma^*\gamma^*\to\pi^0 e^+ e^-$, the leading order
$A$ amplitude only contributes to $F_S$, and the $B$ amplitude
contributes to both $F_S$ and $F_V$ \cite{EPR88, DG95, HS93,
DHV87, FR89}. On the other hand, chiral symmetry enforces that the
scalar form factor $F_S$ be proportional to $m_e$, thus one can
find from Eqs. (\ref{rate1}) and (\ref{AFB1}) that the
electron-positron energy asymmetry ${\cal A}$ due to the
interference between $F_S$ and $F_V$ is suppressed by $m_e^2$,
which is expected to be negligible, as in the case of
$K^+\to\pi^+e^+e^-$ \cite{CQ03, Gao03}. Therefore in the following
we are only concerned about contributions from the $B$ amplitude
to the vector form factor $F_V$, which are given
by\footnote{Actually $O(p^6)$ unitarity corrections to
$K_L\to\pi^0\gamma\gamma$ will induce a dependence on $y^2$ in the
$A$ amplitude, which can also contribute to $F_V$ of $K_L\to\pi^0
e^+e^-$; however, as pointed out in Ref. \cite{BDI03}, this is
numerically rather suppressed \cite{CDM93, CEP93}.}
 \beqn\label{B2gammatoee} {\cal
M}(K_L\to\pi^0\gamma^*\gamma^*\to\pi^0
e^+e^-)^B=\frac{2iG_8\alpha^2}{m_K^2}\int\frac{d^4q}{(2\pi)^4}
\frac{B(y,z)\bar{u}(p_-)\gamma_\mu
v(p_+)}{q^2(Q-q)^2(p_--q)^2}\nonumber\\
\times\left\{p^\mu[2p\cdot q p_-\cdot Q-2p\cdot p_-q\cdot
Q-p\cdot(p_+-p_-)q^2]\right.\nonumber\\
\left. +q^\mu[2p\cdot q p\cdot(p_+-p_-)-2p\cdot Qp\cdot q+2p\cdot
q p\cdot q ]\right\},
 \eeqn
 where $Q=p_++p_-$, $s=Q^2$, and $q$ is the loop momentum for the internal
 photon. The amplitude $B(y,z)$ (which is actually independent of $y$) including unitarity corrections and
 local contributions at $O(p^6)$ reads \cite{CEP93}
 \beqn\label{Bamp}
B(y,z)=c_2\left\{\frac{4
r_\pi^2}{z}F\left(\frac{z}{r_\pi^2}\right)+\frac{2}{3}\left(10-\frac{z}{r_\pi^2}\right)
\left[\frac{1}{6}+R\left(\frac{z}{r_\pi^2}\right)\right]+\frac{2}{3}\ln{\frac{m_\pi^2}{m_\rho^2}}
\right\}+\beta-8 a_V,
 \eeqn
where $c_2=1.1$ is the coefficient ruling the strength of the
unitarity corrections from $K\to 3 \pi$, $\beta=-0.13$, which is
originated from the $O(p^6)$ local terms except the vector
resonance contribution, and the $a_V$ part characterizes the
$O(p^6)$ local contributions generated by the vector resonance
exchange with $a_V=-0.46\pm 0.05$ fixed in the recent NA48
experiment \cite{NA4802}. The functions $F(z/r_\pi^2)$ and
$R(z/r_\pi^2)$ are generated by $\pi$-loop diagrams, which can be
defined as \beq\label{loop}F(x)=\left\{\begin{array}{cc}
1-\Frac{4}{x}\arcsin^2\left(\Frac{\sqrt{x}}{2}\right) & x\leq
4,\\\\1+\Frac{1}{x}\left(\ln{\Frac{1-\sqrt{1-4/x}}{1+\sqrt{1-4/x}}}+i\pi\right)^2&x\geq
4,\end{array}\right. \eeq and
\beq\label{loop2}R(x)=\left\{\begin{array}{cc}
-\Frac{1}{6}+\Frac{2}{x}-\Frac{2}{x}\sqrt{4/x-1}\arcsin\left(\Frac{\sqrt{x}}{2}\right)
& x\leq 4,\\\\-\Frac{1}{6}+\Frac{2}{x}+\Frac{\sqrt{1-4/x}}{x}
\left(\ln{\Frac{1-\sqrt{1-4/x}}{1+\sqrt{1-4/x}}}+i\pi\right)&x\geq
4.\end{array}\right. \eeq Since the integral in Eq.
(\ref{B2gammatoee}) is logarithmically  divergent, only its
absorptive part contribution can be calculated unambiguously.
Actually for the off-shell photons, the $B(y,z)$
 amplitude corresponding to the on-shell photons,  should be replaced
by  $B[y,z, q^2, (Q-q)^2]$. At present, there is no
model-independent way to obtain the off-shell
$K_L\to\pi^0\gamma^*\gamma^*$ form factor. In analogy with the
analysis of the $K_L\to\gamma^*\gamma^*\to\mu^+\mu^-$
\cite{DIP98}, the authors of Ref. \cite{BDI03} proposed the
following ansatz \beqn\label{ansatz} B[y,z, q^2, (Q-q)^2]=B(y,
z)\times f[q^2,(Q-q)^2]\eeqn with the form factor
\beq\label{fqQ-q} f[q^2, (Q-q)^2]=1+a
\left[\frac{q^2}{q^2-m_V^2}+\frac{(Q-q)^2}{(Q-q)^2-m_V^2}\right]+
b\frac{q^2(Q-q)^2}{(q^2-m_V^2)[(Q-q)^2-m_V^2]}\eeq to obtain the
ultraviolet integral by imposing the condition
\beq\label{condition} 1+2a+b=0. \eeq The parameters $a$ and $b$
are expected to be $O(1)$ by naive dimensional chiral power
counting, and in a special case for $a=-b=-1$, the form factor
(\ref{fqQ-q}) will be identical to the one adopted in Ref.
\cite{DG95} for the $K_L\to\pi^0\gamma^*\gamma^*$ transition.
Neglecting terms which are suppressed by powers of $1/m_V^2$ and
eliminating $b$ by means of Eq. (\ref{condition}), one can get
\beqn\label{B2gammatoee2}{\cal
M}(K_L\to\pi^0\gamma^*\gamma^*\to\pi^0
e^+e^-)^B=F_V^{\gamma\gamma}p^\mu \bar{u}(p_-)\gamma_\mu v(p_+),
\eeqn where \beqn\label{FV2gamma} F_V^{\gamma\gamma}=\frac{G_8
\alpha^2 B}{8\pi^2 m_K^2}p\cdot(p_+-p_-)\left[\frac{2}{3}\left(\ln
\frac{r_V^2}{z}+i\pi\right)-\frac{1}{9}+\frac{4}{3}(1+a)\right].
\eeqn Since $F_V^{\gamma\gamma}$ is proportional to
$p\cdot(p_+-p_-)$, the energy asymmetry ${\cal A}$ can be
generated from the interference between $F_V^{\gamma\gamma}$ and
$(F_V^L)_{\rm CPV}$. Using the relation \beq\label{pkp-p} p\cdot
(p_+-p_-)=-\frac{m_K^2}{2}\beta \lambda^{1/2}(1,z,r_\pi^2)
\cos\theta, \eeq one can get \beq\label{AFB3} {\cal
A}(z)=\frac{m_K^7 \beta^2\lambda^2(1,z,r_\pi^2)(1-
\beta^2/2))}{2^{10} \pi^3}{\rm Re}[\tilde{f}_V^*(F_V^L)_{\rm CPV
}]/(d\Gamma/dz),\eeq where $\tilde{f}_V=-
F_V^{\gamma\gamma}/p\cdot(p_+-p_-)$, and $d\Gamma/dz$ is the
differential decay rate after integrating the angle $\theta$ in
Eq. (\ref{rate1}) (neglecting the contributions from $F_S$, $F_P$
and $F_A$).
% which reads \beqn\label{dwidth}
%\frac{d\Gamma}{dz}=\frac{m_K^5 \beta
%\lambda^{3/2}(1,z,r_\pi^2)}{3\cdot 2^8 \pi^3
%}\left[\left(1+\frac{2r_e^2}{z}\right)|(F_V^L)_{\rm CPV}|^2
%+\frac{m_K^4\beta^2 \lambda(1,z,r_\pi^2)}{20}\left(1+\frac{6
%r_e^2}{z}\right)|\tilde{f}_V|^2\right]. \eeqn

Now the electron-positron energy asymmetry defined in Eq.
(\ref{AFB3}) can be determined up to a free parameter $a$, which
is $O(1)$ from the naive dimensional chiral power counting,
however, cannot be fixed from the present theoretical and
experimental studies.\footnote{As pointed out in Ref.
\cite{BDI03}, one can expect to obtain some constraints on this
parameter from the observation of $K_L\to\pi^0
\ell^+\ell^-\gamma$; however, the present experimental data is not
accurate enough to extract any significant constraints
\cite{DG97}.}  For the $O(1)$ values of $a$, the differential
asymmetry ${\cal A}(z)$ can be $O(10^{-1})$, which has been
plotted in Fig. 1 with $a=-1$, $0$, $+1$ and $a_S$ and $a_V$
fixed, respectively.  It is found that the order of magnitude of
the ${\cal A}(z)$ in the present paper is consistent with the one
given in Ref. \cite{DG95}; however, the distribution of the
differential asymmetry against $z$ is not. We would like to give
some remarks on this point here. First, we use $a_V=-0.46$ fixed
in the new NA48 experiment \cite{NA4802} instead of $a_V=-0.96$ in
\cite{DG95}; the vector form factor of $K_S\to\pi^0 e^+e^-$ in Eq.
(\ref{FVS}) has been evaluated up to $O(p^6)$  in chiral
perturbation theory while only the leading order contribution was
considered in \cite{DG95}. Second,  the vector form factor of
$K_L\to\pi^0 e^+ e^-$ via the two-photon intermediate state in Eq.
(\ref{FV2gamma}), which was obtained in \cite{BDI03}, is
inconsistent with the one in \cite{DG95} for the parameter $a=-1$,
where the form factor (\ref{fqQ-q}) is identical to the one
adopted in \cite{DG95}. As pointed out by the authors of
\cite{BDI03}, the main difference between their result and the one
in \cite{DG95} is that they did not find any singularity in the
limit of $m_e\to 0$, and the lacking of this kind of singularity
has been already noticed in \cite{FR89}.

\begin{figure}[t]
\begin{center}
\includegraphics[width=13cm,height=10cm]{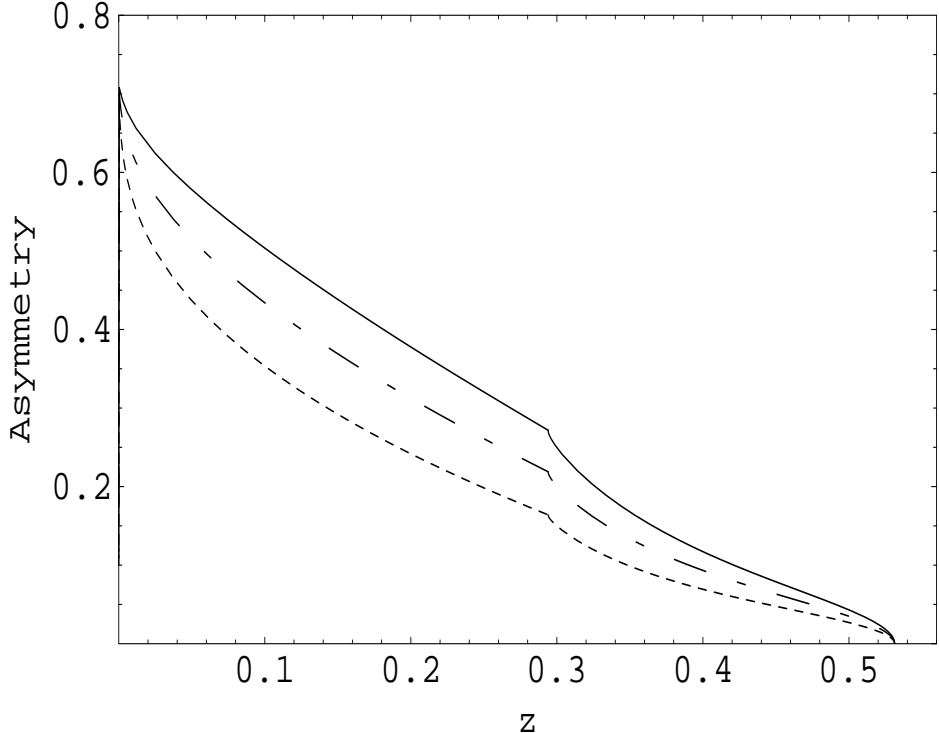}
\end{center}
\caption{ The differential electron-positron energy asymmetry
${\cal A}(z)$ as a function of $z$ with $|a_S|=1.06$ and
$a_V=-0.46$. The full line is for $a=1$, the dotted-dashed line
for $a=0$, and the dashed line for $a=-1$.}
\end{figure}

\begin{figure}[t]
\begin{center}
\includegraphics[width=13cm,height=10cm]{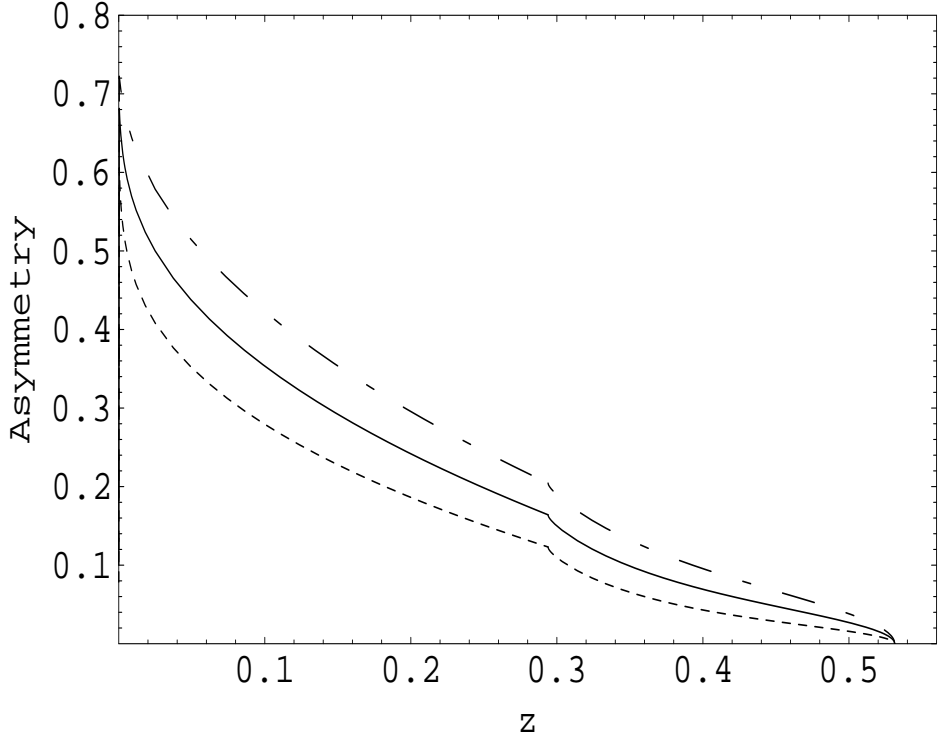}
\end{center}
\caption{ The differential electron-positron energy asymmetry
${\cal A}(z)$ as a function of $z$ with $|a_S|=1.06$ and $a=-1$.
The full line is for $a_V=-0.46$, the dotted-dashed line for
$a_V=-0.51$, and the dashed line for $a_V=-0.41$.}
\end{figure}

\begin{figure}[t]
\begin{center}
\includegraphics[width=13cm,height=10cm]{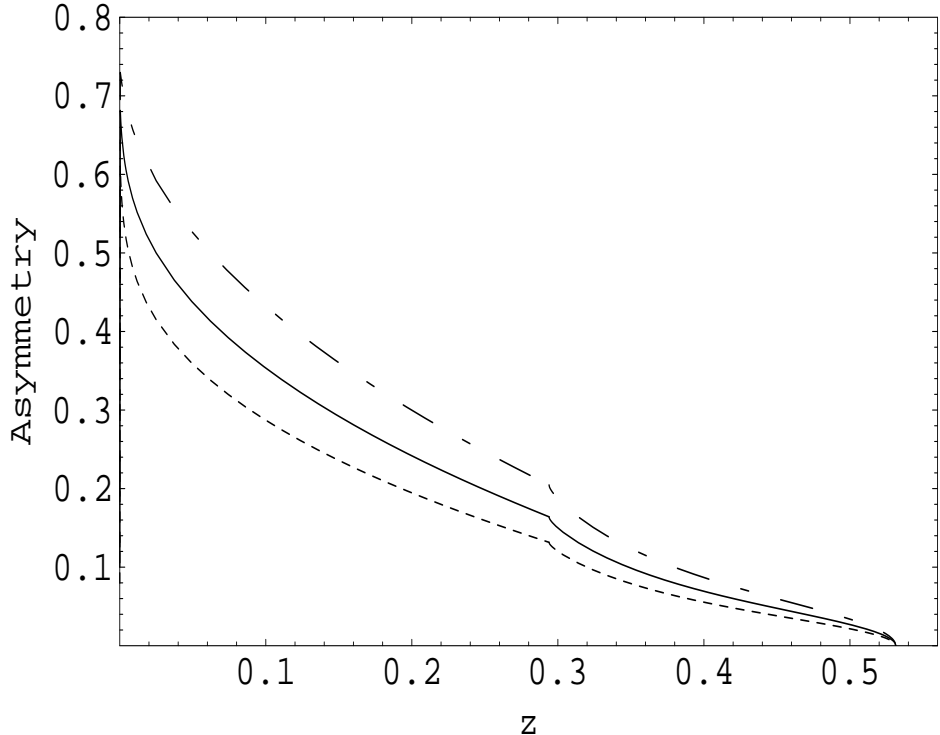}
\end{center}
\caption{ The differential electron-positron energy asymmetry
${\cal A}(z)$ as a function of $z$ with $a_V=-0.46$ and $a=-1$.
The full line is for $|a_S|=1.06$, the dotted-dashed line for
$|a_S|=0.85$, and the dashed line for $|a_S|=1.32$. }
\end{figure}

\begin{figure}[t]
\begin{center}
\includegraphics[width=13cm,height=10cm]{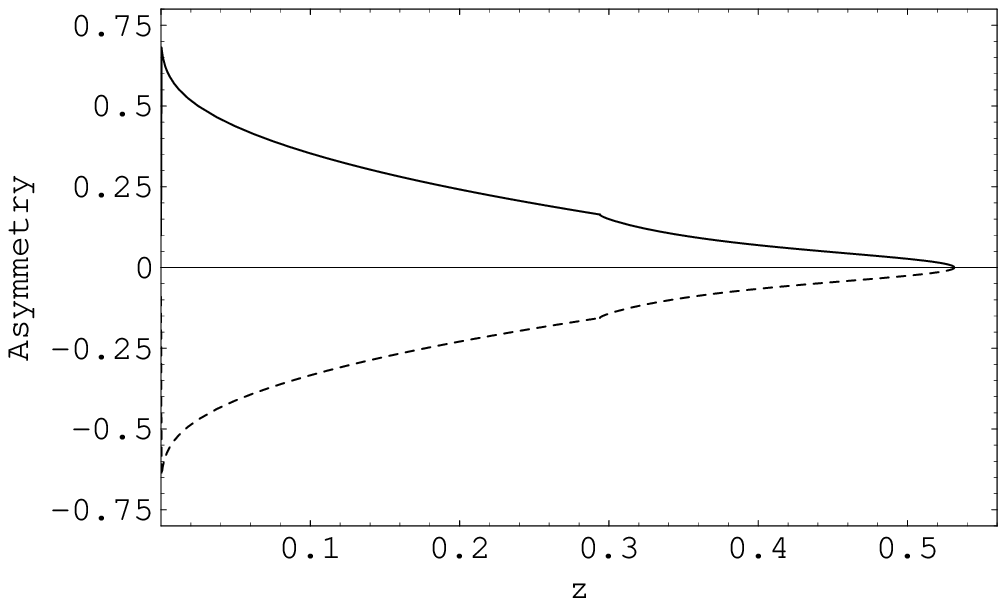}
\end{center}
\caption{ The differential electron-positron asymmetry ${\cal
A}(z)$ as a function of $z$ with $a=-1$, $|a_S|=1.06$, and
$a_V=-0.46$. The full line is for the same sign of  $a_S$ and
$G_8$, and the dashed line for the different sign of $a_S$ and
$G_8$. }
\end{figure}

The relevant contribution from the direct CP-violating component
to the decay rate of $K_L\to\pi^0 e^+ e^-$ is mainly through the
interference between it and the indirect CP-violating one,
therefore it is easily understood that the asymmetry ${\cal A}(z)$
does not depend significantly on the direct CP violation. The main
uncertainty of the asymmetry, besides the one encoded in the
parameter $a$, will come from the indirect CP-violating
contribution due to $K^0-\overline{K^0}$ mixing and the value of
$a_V$. By taking into account the present experimental errors, we
have shown the sensitivity of ${\cal A}(z)$ to $a_V$ and $a_S$ in
Fig. 2 and Fig. 3, respectively, with the fixed parameter $a=-1$.
The similar distributions of the asymmetry as those in Figs. 2 and
3 can be obtained for $a=0$ and $a=1$.

Note that in plotting Figs. 1, 2, and 3, we are not concerned
about the relative sign of $a_S$ and $G_8$, which will be
sensitive to the sign of the asymmetry ${\cal A}(z)$. The new NA48
experimental study of $K_S\to\pi^0 e^+ e^-$ \cite{NA4803} has only
given the absolute value of $a_S$, the sign of $a_S$, which
however cannot be fixed from this measurement, is important to get
the constructive or destructive interference between direct and
indirect CP-violating components when computing the SM prediction
to the total decay rate of $K_L\to\pi^0 e^+e^-$. Theoretically,
the authors of Ref. \cite{BDI03}, using $\Delta I=1/2$ isospin
relation plus VMD argument, have established the sign of $a_S$ in
terms of the sign of $G_8$. Since the indirect CP-violating
amplitude is dominant in this transition, it is not surprising
from Eq. (\ref{AFB3}) that the different relative sign of $a_S$
and $G_8$ will lead to the differential asymmetries with different
sign, which has been shown in Fig. 4. As an example, we take the
parameter $a=-1$ in plotting Fig. 4, and the same conclusions can
be reached for $a=0$ and $a=1$. Thus the observation of the
electron-positron energy asymmetry ${\cal A}(z)$ in $K_{L}\to\pi^0
e^+ e^-$ in future experiments may also be useful to determine the
relative sign of $a_S$ and $G_8$.

We have reexamined the SM contribution to the electron-positron
energy asymmetry in $K_L\to\pi^0 e^+ e^-$, induced by the
long-distance transition from the two-photon intermediate state,
$K_L\to\pi^0\gamma^*\gamma^*\to \pi^0 e^+e^-$. The present
analysis shows that the differential asymmetry ${\cal A}(z)$ can
be $O(10^{-1})$, which implies that more than $10^2$ events would
give the signal of this asymmetry. If the experimental challenges
posed by the so-called Greenlee background \cite{Greenlee} can be
overcome in observing this process, one can therefore expect to
get some interesting information on quark flavor physics by
studying the electron-positron energy asymmetry in the
$K_L\to\pi^0 e^+ e^-$ decay.

The electron-positron energy asymmetry ${\cal A}(z)$ in
$K_S\to\pi^0 e^+ e^-$ can be discussed in analogy with the above
analysis of the decay $K_L\to\pi^0 e^+e^-$. Because the
$K_S\to\pi^0 e^+e^- $ amplitude is dominated by the long-distance
transition via one-photon exchange, and its form factors $F_S$ and
$F_V$ receive no contribution from the two-photon intermediate
state in the limit of CP invariance, ${\cal A}(z)$ in this mode is
expected to be vanishingly small in the SM. Thus the measurement
of the electron-positron energy asymmetry in the $K_S\to\pi^0
e^+e^-$ decay might be very interesting to probe new physics
scenarios beyond the SM.

 \vspace{0.5cm}
\section*{Acknowledgements}
The author wishes to thank G. D'Ambrosio for reading the
manuscript. This work was supported in part by the NSF of China
under Grant No. 10275059.

\end{document}